\documentclass[conference]{IEEEtran}
\usepackage{cite}
\usepackage{amsmath,amssymb,amsfonts}
\usepackage{algorithmic}
\usepackage{graphicx}
\usepackage{textcomp}
\usepackage{xcolor}
\def\BibTeX{{\rm B\kern-.05em{\sc i\kern-.025em b}\kern-.08em  T\kern-.1667em\lower.7ex\hbox{E}\kern-.125emX}}
\usepackage{adjustbox}
\usepackage{listings}
\usepackage{float}
\usepackage{cite}
\usepackage{amsmath,amssymb,amsfonts}
\usepackage{algorithmic}
\usepackage{graphicx}
\usepackage{textcomp}
\usepackage{xcolor,,colortbl}
\usepackage{soul,color}
\usepackage{subcaption}
\usepackage{multirow}
\usepackage{makecell}
\usepackage{booktabs}
\usepackage{tcolorbox}
\usepackage[linesnumbered,ruled]{algorithm2e}
\usepackage{booktabs}
\DeclareCaptionLabelFormat{andtable}{#1~#2  \&  \tablename~\thetable}
\newcommand{\revision}[1]{\textcolor{black}{#1}}
\usepackage[normalem]{ulem}

\usepackage[switch]{lineno} 
\usepackage{colortbl}
\usepackage{fancyhdr}
\usepackage{lastpage}
 
\usepackage[textsize=tiny]{todonotes}
\usepackage{listings}
\usepackage{xcolor}

\definecolor{codegreen}{rgb}{0,0.6,0}
\definecolor{codegray}{rgb}{0.5,0.5,0.5}
\definecolor{codepurple}{rgb}{0.58,0,0.82}
\definecolor{backcolour}{rgb}{0.95,0.95,0.92}

\begin{document}

\title{A Study of Checkpointing in Large Scale Training of Deep Neural Networks}

% \author{\IEEEauthorblockN{Elvis Rojas}
% \IEEEauthorblockA{\textit{Costa Rica Institute of Technology} \\
% \textit{National University of Costa Rica}\\erojas@una.ac.cr}
% \and
% \IEEEauthorblockN{Albert Njoroge Kahira
% \thanks{Albert is with the Barcelona Supercomputing}}
% \IEEEauthorblockA{\textit{Barcelona Supercomputing Center} \\
% \textit{Universitat Politècnica de Catalunya}\\
% albert.kahira@bsc.es}
% \and
% \IEEEauthorblockN{Esteban Meneses}
% \IEEEauthorblockA{\textit{Costa Rica National High Technology Center} \\
% \textit{Costa Rica Institute of Technology}\\
% emeneses@cenat.ac.cr}
% \and
% \IEEEauthorblockN{Leonardo Bautista Gomez}
% \IEEEauthorblockA{\textit{Barcelona Supercomputing Center} \\
% leonardo.bautista@bsc.es}
% \and
% \IEEEauthorblockN{Rosa M Badia}
% \IEEEauthorblockA{\textit{Barcelona Supercomputing Center} \\
% rosa.m.badia@bsc.es}
% }

\author{
    \IEEEauthorblockN{Elvis Rojas\IEEEauthorrefmark{1}\IEEEauthorrefmark{2}, Albert Njoroge Kahira\IEEEauthorrefmark{3}\IEEEauthorrefmark{5}, Esteban Meneses\IEEEauthorrefmark{1}\IEEEauthorrefmark{4}, Leonardo Bautista Gomez\IEEEauthorrefmark{3}, Rosa M Badia\IEEEauthorrefmark{3}\IEEEauthorrefmark{5} }
    \IEEEauthorblockA{\IEEEauthorrefmark{1}Costa Rica Institute of Technology, 
    \IEEEauthorrefmark{2}National University of Costa Rica,
    \IEEEauthorrefmark{5}Universitat Politècnica de Catalunya,\\
    \IEEEauthorrefmark{4}Costa Rica National High Technology Center ,
    \IEEEauthorrefmark{3}Barcelona Supercomputing Center\\
    erojas@una.ac.cr, albert.kahira@bsc.es, emeneses@cenat.ac.cr,\{leonardo.bautista, rosa.m.badia\}@bsc.es
    }
    %\IEEEauthorblockA{\IEEEauthorrefmark{2}National University of %Costa Rica
    %\\erojas@una.ac.cr}
    % \IEEEauthorblockA{}
    %\IEEEauthorblockA{
    %\\\{albert.kahira, leonardo.bautista,rosa.m.badia\}@bsc.es}
    %\IEEEauthorblockA{\IEEEauthorrefmark{4}Costa Rica National High Technology Center
    %\\emeneses@cenat.ac.cr}
}

% \author{Elvis Rojas, Albert Njoroge Kahira, Esteban Meneses, Leonardo Bautista Gomez, Rosa M Badia}
% \IEEEcompsocitemizethanks{
% \IEEEcompsocthanksitem Elvis Rojas, Esteban Meneses,  are with Costa Rica National High Technology Center 
% \protect
% E-mail: \{nguyen.truong, mohamed.attia, takano-ryousei\}@aist.go.jp
% \IEEEcompsocthanksitem Albert Kahira, Leonardo Bautista Gomez, Rosa M Badia are with the Barcelona Supercomputing Center, Spain.\protect \
% E-mail:\{albert.kahira, leonardo.bautista,rosa.m.badia\}@bsc.es}
% %\thanks{* The first two authors contributed equally to this work.}}

\maketitle
\begin{abstract}
Deep learning (DL) applications are increasingly being deployed on HPC systems to leverage the massive parallelism and computing power of those systems. While significant effort has been put to facilitate distributed training by DL frameworks, fault tolerance has been largely ignored. Checkpoint-restart is a common fault tolerance technique in HPC workloads.  In this work, we examine the checkpointing implementation of popular DL platforms. We perform experiments with three state-of-the-art DL frameworks common in HPC (Chainer, PyTorch, and TensorFlow). We evaluate the computational cost of checkpointing, file formats and file sizes, the impact of scale, and deterministic checkpointing. Our evaluation shows some critical differences in checkpoint mechanisms and exposes several bottlenecks in existing checkpointing implementations. We provide discussion points that can aid users in selecting a fault-tolerant framework to use in HPC. We also provide take-away points that framework developers can use to facilitate better checkpointing of DL workloads in HPC.

\end{abstract}
\makeatletter

\def\blfootnote{\xdef\@thefnmark{}\@footnotetext}
\makeatother
\section{Introduction}\label{sec:intro}

Machine Learning (ML) applications have grown in the last decade, and this trend is likely to continue. Research in ML has also grown significantly due to several key factors, such as the enormous amount of data available and significant developments in hardware that has led to increased computing power. In particular, Deep Learning (DL) has emerged as the go-to method for many applications, ranging from computer vision, natural language processing to autonomous driving.

Recent trends in DL include huge models with millions of parameters that require days or even months to train even on high-end hardware such as GPUs. These trends have led to the proliferation of DL applications in High-Performance Computing (HPC) systems which leverage parallelism to provide much-needed computing power. In fact, in the last few years, we have seen HPC systems such as  AI Bridging Cloud Infrastructure (ABCI) \cite{abci}, Summit \cite{summit}, Sierra \cite{Sierra} built specifically for DL workloads.
The proliferation of DL workloads in HPC means that DL applications now face the challenges faced by other HPC applications, such as scalability, IO contention, network congestion and fault tolerance. While extensive research has been done to address most of these issues in many HPC workloads, not much has been done for DL workloads. For instance, not many frameworks are ideal for distributed training in supercomputers. Hence, scaling DL training in distributed clusters often requires significant engineering effort~\cite{Amatya}. To solve this problem, the DL community has created tools such as Horovod~\cite{sergeev2018horovod} that support distributed training on top of existing frameworks.

While distributed training in  HPC clusters can reduce weeks of training to days or even hours, these distributed architectures are susceptible to unrecoverable hardware and software failures that can ruin days or even weeks of training time. To mitigate these failures, most DL frameworks implement checkpointing as fault tolerance mechanisms to save the training state at a certain point. In case of failure,  training is restored from the last valid checkpoint. However, checkpointing in DL is not just a defence mechanism against failures. Transfer learning, a common technique in DL relies on checkpointing. %A model is trained on a different data set, saved, and finally fine-tuned with the target application data. This significantly saves on training time and resources. 
Recently, gradient checkpointing has been used to overcome memory constraints by fitting large models in GPU memory and the cost of increased computational cost. Checkpointing is, therefore, a fundamental component of training DL models.% and even more important when DL such models are trained in HPC.

Despite being such a critical component of training DL models, checkpointing has largely been ignored by the DL community and checkpoint implementations of DL frameworks are not ideal for HPC. Related work (See section~\ref{sec:related_work}) shows that there is a need for advanced checkpointing mechanisms for DL workloads, especially in HPC. However, to improve existing mechanisms or create new ones, an in-depth study and evaluation of checkpointing in large scale training of DNNs is required.   While checkpointing provides a way to restart applications on failure, it is important to ensure that on a restart, the behaviour of such applications is maintained.  This ensures reproducibility and validation. %The reproducibility of experiments is an essential element of scientific research that permits scientific integrity. 
Furthermore, it is important for researchers who want to analyze the behaviour of DL applications under particular circumstances~\cite{gundersen}.

In this paper, we provide detailed explanations of how checkpoint mechanisms work to illustrate the design decisions involved. We then evaluate checkpointing using representative DL models with three state-of-the-art DL frameworks common in HPC: Chainer, PyTorch, and TensorFlow. \revision{ We use Cifar \cite{cifar}  dataset because it provides an adequate number of images that allow DL training to finish within acceptable execution times therefore ideal for experimentation. In addition, this dataset allowed us to properly analyze the performance of distributed training when scaling on GPUs.}  We further study deterministic behaviours in these frameworks to validate reproducibility when checkpointing. In summary, the contributions of this paper can be summarized as follows. 

\begin{enumerate}
\item We  explore, explain, and compare checkpoint mechanisms of distributed computing DL frameworks.

\item We perform a set of  experiments to  measure and evaluate checkpoint overhead at different scales.

\item We study the deterministic execution of DNN training. Such experiment is fundamental for reproducibility. 
\end{enumerate}

The remainder of this paper is organized as follows. Section~\ref{sec:background} gives a detailed background of distributed DL and different frameworks used in DL. Section~\ref{sec:methodology} explains our experimentation methodology. Section~\ref{sec:evaluation} shows our evaluation and results. Section~\ref{sec:discussion} discusses important points for future improvement. Section~\ref{sec:related_work} presents related work and Section~\ref{sec:conclusion} concludes the paper.

\section{Related Work}\label{sec:related_work}

As mentioned earlier, checkpointing in DL has been largely ignored by both  the DL and HPC community. However, recent works show a growing interests. Early work by Vinay et al \cite{Amatya} addressed the requirements of MPI for designing fault tolerant DL applications, more specifically checkpoint-restart. Reagen et al \cite{Ares} proposed a framework for quantifying the resilience DNNs. Qiao et. al.~\cite{qiao} also quantify the impact of faults on interactive-convergent algorithms and they proposed strategies based on checkpoint to tolerate faults in distributed training. Most recently Nicolae et al ~\cite{nicolae} proposed a checkpoint framework based on  VeloC~\cite{veloc} that acts as a bridge to advanced techniques used in multi-level HPC checkpointing, taking advantage of I/O patterns, layer-wise parallelism and the properties of the synchronous data parallel training.  Besides fault tolerance, ~\cite{beaumont} checkpoints are used to decrease memory usage when training deep neural networks with the back-propagation algorithm.

On deterministic DL and reproducibility, recent work include \cite{islam,Icke,Silver,Ilievski}.  These studies use deterministic approaches to generate benchmarks or algorithms to increase the performance or reliability of machine learning related applications.  Also, in~\cite{islam} one of the main goals is the reproducibility of the results.  Another study~\cite{nagarajan} is focused on analysis of positive impacts of deterministic implementations in training.  They identify the sources of  non-determinism in DL applications and some experimental conditions that form a gap between deterministic implementations and replicability.

\section{Background}\label{sec:background}

The concept of DL originated from the study of artificial neural networks (ANN). %, which during the last decades has become a very active area of research. 
Depending on the problem to be solved, the process of training an ANN can take a large number of causal events in the computational stages generating performance problems in backpropagation and overfitting. %The birth of DL is marked in 2006, when a new training method called layer-wise-greedy learning~\cite{Hinton} was proposed with the basic idea that unsupervised learning could be effectively pre-trained one layer at a time.  
A DNN is an ANN with multiple layers between input and output. Based on this, a DL algorithm can be defined as a hierarchical architecture with many layers each consisting of non-linear information processing units~\cite{LIU201711, Schmidhuber_2015}.
%DL applications have increased considerably nowadays, which has promoted that DNN models are becoming more complex while increasing the amount of data they process. The previous premise together with 
The emergence of large distributed computing systems have led to the evolution of DNNs to adapt to the new advantages that distributed training parallelization provides.
%\subsection{Distributed Training }\label{sec:training_parallel}
%\subsubsection{Data and model parallelization}
Distributed training means that it is necessary to parallelize the training phase of a DL model. There are two parallelization strategies.  The most common strategy is based on the parallelization of the  training data~\cite{Krizhevsky} and another one in the parallelization of the  DL model~\cite{harlap,chen}. In this work we focus the data parallelization strategy.

%In data parallelization each GPU holds a copy of the entire network and each GPU processes a different portion of the  training data.   Each GPU computes its own gradients and loss with respect to the data it possesses.  Before finish each epoch the gradients and loss are averaged among all GPUs through a collective operation like \texttt{MPI\_Allreduce}. 
%Data parallelism suffers from intra-GPU communication overheads, since each individual GPU needs to be synchronized.  These communication overheads is proportional to the model size and becomes a scalability bottleneck in large scale distributed training~\cite{Dean,Aji}. 
%Model parallelism divide the model into disjoint subsets and each subset is assigned on a dedicated GPU.  Each GPU is liable for the updates of the assigned model layers.   Model parallelization is appropriate when the model is excessively large to be fitted into a single GPU due to the memory capacity.  Nevertheless, split the model into subsets is not an easy task because it could generate load imbalance issues limiting the scaling~\cite{harlap,chen}.

\subsection{Distributed Data Training Mechanisms}\label{sec:frameworks}
Almost all major DL frameworks provide some sort of support for distributed training of DNNs.  We use 3 different DL frameworks to conduct distributed training:  Chainer~\cite{tokui,tokui2015}, PyTorch~\cite{paszke}, and TensorFlow~\cite{tensorflow1, tensorflow2}.  All these DL frameworks provide support for GPUs and allow the use of libraries like CUDA, CuDNN and NCCL.

The current trend of DNN applications to accelerate training is the use of HPC systems. Systems that have architectures with more than one GPU per node are particularly targeted. This has lead DL frameworks to evolve and implement native mechanisms to support distributed training.
%However, there are open source libraries from other developers that are used to carry out distributed training with the advantage of easy implementation or without making substantial changes to already developed DNN applications.
In this work, distributed training is implemented in two ways: $i)$~with the native resources of the DL framework (we use this approach with Chainer and PyTorch), $ii)$~with external libraries to the DL framework (we use Horovod with PyTorch and TensorFlow).
Each of the DL frameworks implements distributed training differently. In the case of Chainer we used the \texttt{ChainerMN} library~\cite{tokui}. With PyTorch we use \texttt{DistributedDataParallel} (DDP)~\cite{dataparallel}.  When Horovod works with PyTorch or TensorFlow it wraps the optimizer with \texttt{hvd.DistributedOptimizer()}. Then, it uses all-reduce operations to combine gradient values before applying gradients to the model weights.  Horovod also uses other instructions to broadcast model-state parameters and the optimizer state from root rank to all other processes to ensure a consistent initialization~\cite{hvd_api}.

\subsection{Deterministic Behavior of DNN Training}\label{sec:deterministic}

There is an intrinsic randomness in the results of the training even with the same infrastructure (hardware, framework versions, etc.) and model configurations.  This can raise reasonable questions about the reliability of the results after restarting a saved checkpoint.  So, in order to validate the appropriate functioning of the checkpoints we want to obtain a deterministic behavior after restart. This means that accuracy and loss values after restart are exactly the same as for a complete training (we call this \emph{deterministic restart}). %With this we make sure that the checkpoint mechanism allows us to continue with a correct execution.  %The main changes made in the DL applications to obtain deterministic results in their respective training are detailed below.

To remove non-deterministic behavior from the three DL frameworks, we basically added instructions that disable the randomness of the internal processes of the DL framework, and of the libraries external to the DL framework (e.g. CuDNN, NumPy, or CuPy). It is necessary to set the seed of the instructions that generate randomness to an equal and constant value when starting the DL applications.  With this, it is possible to generate identical random sequences that are used by the internal processes of the DL frameworks.  Because distributed trainings use GPUs,  some instructions to control randomness are related to CUDA and cuDNN~\cite{cudnn} libraries.  Also, it was not possible to implement deterministic restart in both Chainer and TensorFlow, since training information is encapsulated into a single container. Such design decision makes it highly cumbersome to adapt either Chainer or TensorFlow for a deterministic restart (See Section~\ref{sec:deterministic_results}).

\subsection{Checkpointing in DNN Training }\label{sec:ft_dnn}

%One of the main goals of this research is to analyze the fault tolerance of distributed DL models. We decided to analyze the behavior of distributed training after catastrophic faults in which execution is completely interrupted.  For this, checkpoint mechanisms are implemented, which are provided by the same DL frameworks. %With these, it is possible to restart distributed training from a specific point. In our case a specific epoch.

%\subsubsection{Checkpoint implementation }\label{sec:Checkpoint implementation}
%\todoLeo{No point on having III.A. if there is no III.B.}

In the DL frameworks used, the implementation of the checkpoint mechanisms is straightforward. However, the implementation is not automatic. % and depending on the DL framework it may involve more modifications in the code to incorporate them. 
In addition, DL frameworks allow to select the state data type  to save in the checkpoint and the output file type.
The implementation of checkpoints varies according to the DL framework. However, these share the possibility of saving in the same HDF5 format. PyTorch  provides the \texttt{torch.save()} and \texttt{torch.load()} functions to save the current state (training state checkpoint) and to load a saved checkpoint, respectively.  The serialization process is carried out with the Pickle library~\cite{pickle}.

With Chainer we implement checkpoints based on the extension \texttt{snapshot()}. %This extension allows serialization of a trainer object to save it to an output file. % By using this extension, it is possible to resume a training from a saved state. The extension uses as one of its parameters a trigger that tells Chainer when to perform the checkpoint, e.g. \texttt{trigger \= (1, 'epoch')} indicating that the action is executed on each epoch. 
%This extension must be added as an extension to the Chainer trainer through the \texttt{chainer.training.Trainer.extend()} function.  
In order to load a checkpoint, it is necessary to use the \texttt{load\_npz()} function before the execution of the training begins. These functions are used to serialize and store in NPZ (Numpy’s compressed array format) format. Chainer also supports native HDF5 format storage.

TensorFlow uses a callback called \texttt{ModelCheckpoint()} to implement the checkpoints.  Through this callback it is possible to configure parameters to manipulate the checkpoint process. %such as the frequency of the checkpoints (\texttt{save\_freq}) or if the complete model is stored or only its weights (\texttt{save\_weights\_only}).  Furthermore, this callback has an interesting parameter that allows to save the best model (\texttt{save\_best\_only}) according to a monitoring (\texttt{monitor}), either of the accuracy or the loss.
The serialization process can result in two types of file format, one native to TensorFlow and one HDF5. %To get the files in HDF5 format TensorFlow uses the H5PY library as part of the serialization process.  %This differs from PyTorch in that the H5PY library is not implemented by default. 
TensorFlow by default saves files in their native format. % and to indicate that a file in HDF5 format is needed only requires setting the file extension of the output file to $.h5$.  In case the callback input argument is \texttt{save\_weights\_only = false} TensorFlow will create a checkpoint of the entire model saving the model architecture (to allow to re-instantiate the model), the model weights, and the state of the optimizer to allow you to restart a training exactly where it left off.  
%On the other hand, 
The deserialization process does not require callbacks. TensorFlow can load weights or the entire model.
% \texttt{tf.keras.models.load\_weights()} function or the entire saved model with the \texttt{tf.keras.models.load\_model()} function. % It is clear that the function \texttt{tf.keras.models.load\_weights()} can only be used if the parameter \texttt{save\_weights\_only = true} in the callback that generated the checkpoint.  The format of the file to load is irrelevant to either of the two functions and again the file extension allows TensorFlow to determine and execute the correct functions.

%\begin{figure}[t]    \centerline{\includegraphics[scale=0.3]{figures/chekpoint_pytorch.png}}
 %   \caption{General diagram of the checkpoint and restart process in PyTorch.}
 %   \label{fig:check_pytorch}
%\end{figure}

\begin{figure}[t]    \centerline{\includegraphics[scale=0.28]{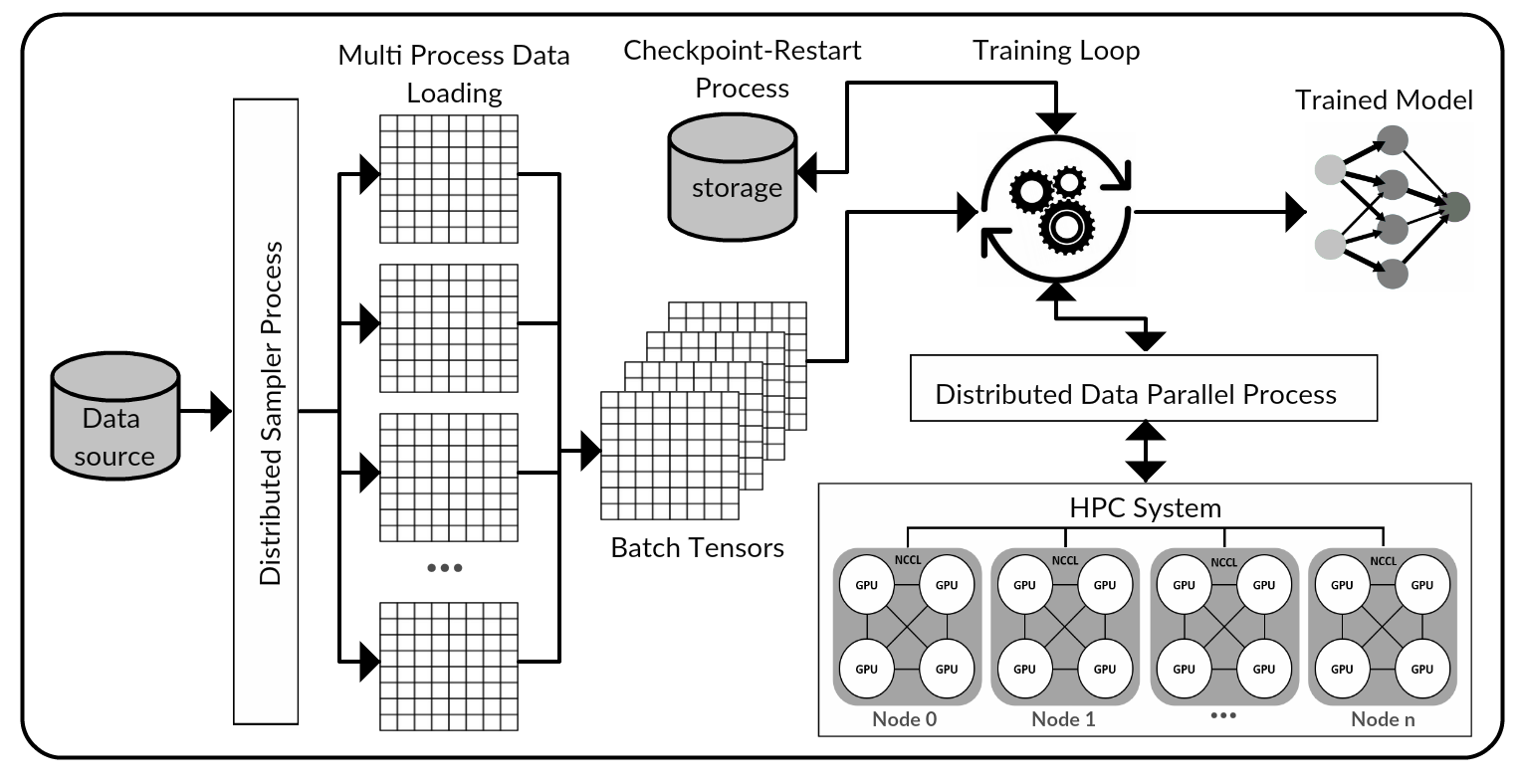}}
    \caption{Distributed Training Process.}
    \label{fig:distributed_training}
\end{figure}

%\end{table*}
%\hfill
%\begin{minipage}{.4\textwidth}
%%\centering
%\footnotesize
%\setlength{\tabcolsep}{1.5 pt}
%\begin{tabular}{c|c|cc|cc}\hline
%\toprule[0.5pt]
%\midrule[0.3pt]

% \multirow{2}{*}{\textbf{Framework}}& & \multicolumn{2}{c|}{\textbf{Checkpoint}}  & \multicolumn{2}{c}{\textbf{Restart}}  
%& &  & \textbf{Updater}  & \textbf{Model} & \textbf{Updater} & \textbf{Model} 
%\\\hline
% \multirow{3}{*}{Chainer} &AVG	&	0.27	&	0.46	&	0.22	&	0.41 \\%\hline
%&STD	&	0.011	&	0.015	&	0.014	&	0.035	\\%\hline
%&CV	&	4.1	    &	3.25	&	6.48	&	8.51	%\\\hline

%\\\hline
% \multirow{3}{*}{PyTorch} &AVG	&	\multicolumn{2}{c|}{0.15}			&	\multicolumn{2}{c}{0.13}	  \\%\hline
%&STD	        &   \multicolumn{2}{c|}{0.013}			&	\multicolumn{2}{c}{0.015}			\\%\hline
%&CV	&	\multicolumn{2}{c|}{8.58}	     		& \multicolumn{2}{c}{11.15	}	\\\hline

% \multirow{3}{*}{TensorFlow}&AVG	& \multicolumn{2}{c|}{0.45}				& \multicolumn{2}{c}{0.33	}  \\%\hline
%&STD	& \multicolumn{2}{c|}{0.034}			& \multicolumn{2}{c}{0.013	}		\\%\hline
%&CV	&	 \multicolumn{2}{c|}{7.63 }  	&	\multicolumn{2}{c}{4.14}	\\\hline
%\toprule[0.5pt]

%\end{tabular}
%\caption{Statistical analysis.}
%\label{tbl:save_load_check_times}
%\end{minipage}

\section{Methodology}\label{sec:methodology}

Fault tolerance in DL has been largely ignored by the community~\cite{nicolae}. Perhaps due to the fact that, traditionally, most of the research was carried out on a single GPU or single node multi GPU environment. Furthermore, most of the DL frameworks are optimised for performance. However in HPC, fault tolerance is a fundamental component because faults and node failures are common with libraries such as FTI~\cite{fti} built specifically to facilitate fault tolerance of applications in HPC. 

Figure~\ref{fig:distributed_training} shows the general structure of a multi-GPU training or distributed training with the implementation of the checkpoint process as a fault tolerance mechanism. In the diagram, the interaction between the checkpoint-restart mechanism and the training cycle is visible. It gives us an idea of the behavior of the checkpoint-restart and its relationship with the performance of distributed training.

We aim to evaluate existing support for fault tolerance in state-of-the-art DL frameworks and their readiness for HPC. This information is important for framework designers and users alike. For framework designers, our aim is to test, evaluate and validate existing implementations in HPC context. For users, our aim is to highlight key differences, similarities and trade offs in choosing frameworks vis-à-vis of fault tolerance. In a nutshell, we evaluate the following: $i)$~The computational cost of checkpointing, $ii)$~Efficiency of inbuilt checkpointing mechanisms in different frameworks, $iii)$~The effect of scale on checkpointing, and $iv)$~The deterministic behaviour in different frameworks. We perform this experiments using common machine learning models and datasets on two state-of-the-art HPC systems; Marenostrum and ABCI.

\textbf{\emph{Power9 Cluster in the Marenostrum Supercomputer}}  is a cluster of 52 nodes and each node is made up of 2 IBM Power9 processors and 4 NVIDIA V100 GPUs.  Power9 processors can reach a frequency of 3 GHz, with 20 cores and 4 threads per core (160 threads per node). Each of the V100 GPUs has 16 GB HBM2 memory, with a performance of 7.8 teraFlops in double precision and 125 teraFlops with the tensor cores~\cite{v100}. This cluster also has a high performance distributed file system (IBM GPFS), which allows access to data from all nodes in the cluster.  Then, parallel applications can simultaneously access data from any node that has the file system.

\textbf{\emph{ABCI Supercomputer}} comprises 1,088 nodes of FUJITSU Server PRIMERGY CX2570 M4. Each compute node  has two Intel Xeon Gold 6148 Processors and four NVIDIA Tesla V100 GPUs (16GB of memory per GPU). The GPUs are connected intra-node to the CPUs by PLX switches and PCIe Gen3 x16 links, and together by NVLink.

Our experiments are largely divided into two groups. The first set of experiments is performed on Marenostrum, while the second on ABCI. Though these two systems share similar hardware characteristics, they are different on scale. As such, we are able to comprehensively evaluate both systems. 

The first set of experiments evaluates the computational cost of checkpointing (Subsection~\ref{sec:comp_cost}), the checkpoint file size and format (Subsection~\ref{sec:checkpoint_size_format}), and deterministic behavior of the checkpoint-restart mechanisms of different frameworks (Subsection~\ref{sec:deterministic_results}).
We use the deterministic behavior to validate the restart functionality after saving a checkpoint and we determined the performance degradation this incurs.

We use Horovod with PyTorch and TensorFlow to implement distributed training.
These experiments were performed at Marenostrum4 supercomputer and up to 32 GPUs were used. Cifar10 was used as the dataset, ResNet50 with batch size of 64. 
\revision{We only ran ResNet50 on Marenostrum4 as we were using this platform to explore checkpoint features. Large-scale experiments used more neural network models.
}

The second set of experiments is to evaluate the effect of scale on checkpointing, the behaviour of different models and how each framework performs at scale. These set of experiments were performed on the ABCI supercomputer using up to 256 GPUs. Our goal is to evaluate the computation cost of checkpointing at such scales and the performance of different frameworks. We use ResNet50 and VGG16 models. These two models have different characteristics that make them interesting to study. VGG16 is a deep and dense model with 138 million parameters. ResNet50 is a residual network with skip connections and 25.5 million parameters. We train both models on the Cifar10 dataset with a batch size of 32 (Subsection~\ref{sec:checkpoint_scale}).
\revision{ Both sets of experiments were run in all cases with 100 training epochs. When training with checkpoint, checkpoint frequency was set to 5 epochs.}

\begin{table}[t]

\begin{minipage}{.48\textwidth}
\centering
\footnotesize
 \caption{Distributed training time for 100 epochs. Columns show the execution times for a training without checkpoint (NoCkpt), with checkpoint (Ckpt), and the percentage of checkpoint overhead ($\Omega$).}
\setlength{\tabcolsep}{3.2 pt}
\begin{tabular}{c|ccc|ccc|ccc}\hline
\toprule[0.5pt]
\midrule[0.3pt]
 & \multicolumn{3}{c|}{\textbf{Chainer}} & \multicolumn{3}{c|}{\textbf{PyTorch}}& \multicolumn{3}{c}{\textbf{TensorFlow}}  \\\hline
\textbf{GPU}  & \textbf{NoCkpt} & \textbf{Ckpt} & $\Omega$ &\textbf{NoCkpt} & \textbf{Ckpt} & $\Omega$ & \textbf{NoCkpt} & \textbf{Ckpt} & $\Omega$ \\\hline
4	&3119	&3240	&3.8&1801& 1826& 1.3 & 1107 &  1124 & 1.5     \\%\hline
8	&1726	&1869   &8.2&885 & 896 & 1.2 & 633  &  648  & 2.3 	  \\%\hline
12	&1283	&1451	&13.1 &601 & 623 & 3.6 & 497  &  504  & 1.5 	   \\%\hline
16	&1013	&1153	&13.7 &454 & 465 & 2.5 & 412  &  420  & 1.9      \\%\hline
20	&862	&1006	&16.7 &374 & 380 & 1.6 & 371  &  373  & 0.3 		\\%\hline
24	&762	&902	&18.4 &310 & 320 & 3.1 & 329  &  351  & 6.8 		\\%\hline
28	&699	&824	&17.9 &288 & 299 & 3.7 & 325  &  329  & 1.4 		\\%\hline
32	&633	&773	&22.1 &278 & 294 & 5.5 & 313  &  322  & 2.9 		\\\hline
\toprule[0.5pt]
\end{tabular}
   
\label{tbl:times}
%\todo{Why is this table in methodology section }
\end{minipage}
\end{table}

\section{Evaluation and Results}\label{sec:evaluation}
%In this section we present our large scale evaluation and the results of our experiments.

\subsection{Computational Cost of Checkpointing}\label{sec:comp_cost}
Saving the state of a DNN consumes a significant amount of time, in particular if it is done at high frequency. Table~\ref{tbl:times} shows the total time to complete a training without checkpointing, with checkpointing, and the percentage of overhead represented by NoCkpt, Ckpt, and $\Omega$, respectively. We performed this experiment for the three DL frameworks, Chainer, PyTorch, and TensorFlow.  For each of the frameworks, distributed training runs were executed with different number of GPUs, ranging from 4 GPUs (1 node) to 32 GPUs (8 nodes). In addition, each of the training runs was repeated 5 times and average time is reported.

As the number of GPUs increases, the performance improves accordingly, both for training with and without checkpoint. PyTorch had the best performing training times with a speedup of 6.4 with 32 GPUs without checkpoint and 6.2 with checkpoint. PyTorch is the framework that presents the best performance when scaling on GPUs. It is also interesting to note that training times with and without checkpoint in TensorFlow had a speedup of 3.5 for both training with and without checkpoint. Chainer obtained the lowest performance in these small-scale experiments. However, Chainer's performance excels at large-scale tests (See Section~\ref{sec:checkpoint_scale}).

Regarding the checkpoint overhead, the lowest overall average overhead is obtained by TensorFlow with 2.3\%, followed by PyTorch with 2.5\%.  Chainer has the highest overall overhead of 14.27\%.  The best performance in TensorFlow can be attributed to the use of the HDF5 file format, because the serialization process of this file format is highly optimized.  Furthermore, it can be seen that Chainer is the only one with a constant increase in overhead, which is proportional to the number of GPUs going from 3.8\% with 4 GPUs to 22.1\% with 32 GPUs. The other DL Frameworks maintain a relatively constant overhead fluctuating in most cases between 1\% and 3\% for PyTorch and 1\% and 2\% for TensorFlow.

Finally, checkpoint mechanisms do not seem to alter the performance profile of the DL frameworks when scaling in small-scale experiments. Although the checkpoint overhead is not negligible, it does not susbstantially affect training performance for PyTorch and TensorFlow. In the case of Chainer, it is necessary to carry out a more detailed review of the checkpoint operation to optimize the mechanisms for the checkpoint process. In this study, the performance of the checkpoint in Chainer was not optimized to be consistent with the other DL frameworks.

\begin{table}[t]
\centering
\footnotesize
%\resizebox{\textwidth}{!}{%
\caption{Comparative of the size and format of the checkpoint between the different DL frameworks.}
\setlength{\tabcolsep}{9.2 pt}
\begin{tabular}{c|c|c|c|c} \hline
\toprule[0.5pt]
\midrule[0.3pt]
 & \textbf{Model} & \textbf{Chainer}  & \textbf{PyTorch} & \textbf{Tensorflow}    \\ \hline

\multirow{2}{*}{Size(MB)} & ResNet50 & 146 & 180 & 181  \\ %\hline
 & VGG16 & 238 &1025  & 257 \\ \hline
\multicolumn{2}{l|}{Format}   & NPZ & Pickle & HDF5  \\ \hline
\toprule[0.5pt]
\end{tabular}%
%}

\label{tbl:check_sizes}
\end{table}

\begin{table*}[ht]
%\begin{adjustbox}{width=\textwidth}
\footnotesize
\caption{Distributed training running times for 100 epochs in ABCI supercomputer for VGG16 and ResNet50 models.}
%\resizebox{\textwidth}{!}{%
\setlength{\tabcolsep}{3.7 pt}
\begin{tabular}{c|ccc|ccc|ccc|ccc|ccc|ccc}
\toprule[0.5pt]
\midrule[0.3pt]
%\hline
 & \multicolumn{9}{c|}{\textbf{RESNET50}} & \multicolumn{9}{c}{\textbf{VGG16}} \\ \hline
 & \multicolumn{3}{c|}{\textbf{Chainer}} & \multicolumn{3}{c|}{\textbf{PyTorch}} & \multicolumn{3}{c|}{\textbf{TensorFlow}} & \multicolumn{3}{c|}{\textbf{Chainer}} & \multicolumn{3}{c|}{\textbf{PyTorch}} & \multicolumn{3}{c}{\textbf{TensorFlow}} \\ \hline
\textbf{GPU} & \textbf{NoCkpt} & \textbf{Ckpt} & \textbf{$\Omega$} & \textbf{NoCkpt} & \textbf{Ckpt} & $\Omega$ & \textbf{NoCkpt} & \textbf{Ckpt} &\textbf{$\Omega$} & \textbf{NoCkpt} & \textbf{Ckpt} &$\Omega$ & \textbf{NoCkpt} & \textbf{Ckpt} &$\Omega$ & \textbf{NoCkpt} & \textbf{Ckpt} & $\Omega$  \\ \hline
4  & 2162& 2338&8.1 & 1738&1801&3.6 &1856&1936&4.3&647&880 &36.1  &1106&1130&2.1 &813&817&0.4 \\ %\hline
8  & 1143& 1295&13.2 & 933 &941 &0.8 &1080&1082&0.1&796&1023&28.5  &1322&1353&2.3 &615&622&1.1 \\ %\hline
16 & 600 & 747 &24.5 & 484 &493 &1.8 &602 &618 &2.6&355&584 &64.5  &721 &750 &4.1 &385&395&2.5 \\ %\hline
32 & 307 & 446 &45.2 & 253 &259 &2.3 &364 &371 &1.9&180&415 &130.5 &370 &402 &8.6 &235&244&3.8 \\ %\hline
64 & 157 & 302 &92.3 & 140 &142 &1.4 &236 &248 &5.1&95 &336 &253.6 &198 &231 &16.6&150&159&6            \\ %\hline
128 &83  & 228 &174.6& 77  &84  &9.0 &174 &186 &6.8&52 &291 &459.6 &113 &141 &24.7&114&128&12.2 \\ %\hline
256 &47  & 190 &304.2& 48  &54  &12.5&149 &157 &5.3&31 &270 &770.9 &70  &98  &40  &99 &110&11.1 \\ \hline
\toprule[0.5pt]
\end{tabular}
%\end{adjustbox}
\label{tab:abci_times}
\end{table*}

\subsection{Checkpoint File Size and Format}\label{sec:checkpoint_size_format}

%Due to the current storage capacities of HPC systems, the size of checkpoint files has not been considered as a key optimization element in DL frameworks.
\revision{ With the increase in complexity of the DL models, the size of the checkpoint files also increases, so the size of the files should be considered as a key element within the optimization of DL frameworks.}
With each framework that is developed, the range of file formats expands. Table~\ref{tbl:check_sizes} shows the formats and sizes of the files saved in each checkpoint for each of the DL frameworks that we used in the experiments. In addition, the table is also divided according to the neural network model that was used. We wanted to determine if the neural network used can influence the size of the resulting checkpoint files.

Although in all frameworks the HDF5 format can be used, we decided to test the checkpoint in the formats that are specific to the implementation of each DL framework. The only one that implements a native file format without the use of third-party libraries is TensorFlow.  In the experiments with TensorFlow we used HDF5 because in the native file format it generates large files which could generate overhead, disadvantaging the performance of TensorFlow when it is compared to other DL frameworks. For reference,  using ResNet50 as model, the checkpoint file in native TensorFlow format reached approximately 427 MB. On the other hand, PyTorch uses Pickle to serialize, so its file format is based on it.  In the case of Chainer the resulting file format is NPZ and is based on the NumPy library. It is interesting to note that the NumPy library internally uses Pickle to implement serialization.

If we compare the file sizes by neural network these do not vary considerably with ResNet50. PyTorch and TensorFlow maintain similar sizes, while with Chainer the file size is approximately 19\% smaller.  It is interesting how well optimized the serialization process is in Chainer, however the price of this optimization is paid with the degradation in performance. On the other hand, with the VGG16 model, all file sizes show a notable increase.  This is to be expected as VGG16 has 138 million parameters \cite{simonyan2014deep} compared to ResNet50's 25 million \cite{zagoruyko2016wide}.  Also, there is a notable increase in file size with PyTorch increasing by 469\% and beating Chainer by 331\% and TensorFlow by 299\%.  This large increase shows that PyTorch's serialization mechanisms may not be optimized for different neural network models, unlike Chainer and TensorFlow in which file sizes maintain proportionality increasing from ResNet50 to VGG16 by 63\% in Chainer and 42\% in TensorFlow.

If we compare the file sizes between the DL frameworks, we can see that there are notable differences that are attributable to the file formats used by each DL framework.  Chainer and Pytorch file formats are based on the Pickle library which might suggest that the file sizes should be similar. However, these two DL frameworks perform serialization differently which can influence the size of the resulting file.  Chainer uses Pickle through NumPy to serialize to NPZ format, and PyTorch uses pickle directly as part of its checkpoint implementation.  
Also, even though the HDF5 file format has optimized compression, it does not show a significant difference with the Chainer file size, so using NPZ-based formats can be a good decision. Taking into account the above argument and if there is no user concern for storage space, any of the 3 file formats performs its function adequately and attention should be directed to the performance of each of these.  However, DL models and DL applications are growing in complexity and size, so in the short term it will be necessary to pay more attention to the serialization and compression processes of the files resulting from the checkpoints.

%For relatively small DNN models, the end users are not concerned about the format or size of the resulting file, because the checkpoint mechanism is used relatively little during training and may not cause storage problems.  However, DL models and DL applications are growing in complexity and size, so in the short term it will be necessary to pay more attention to the serialization and compression processes of the files resulting from the checkpoints.

%--------------------------------- Checkpointing at Scale-----------------------% 

\begin{figure*}[t]
  \begin{center}

    \begin{subfigure}{.49\hsize}
      \includegraphics[width=\hsize]{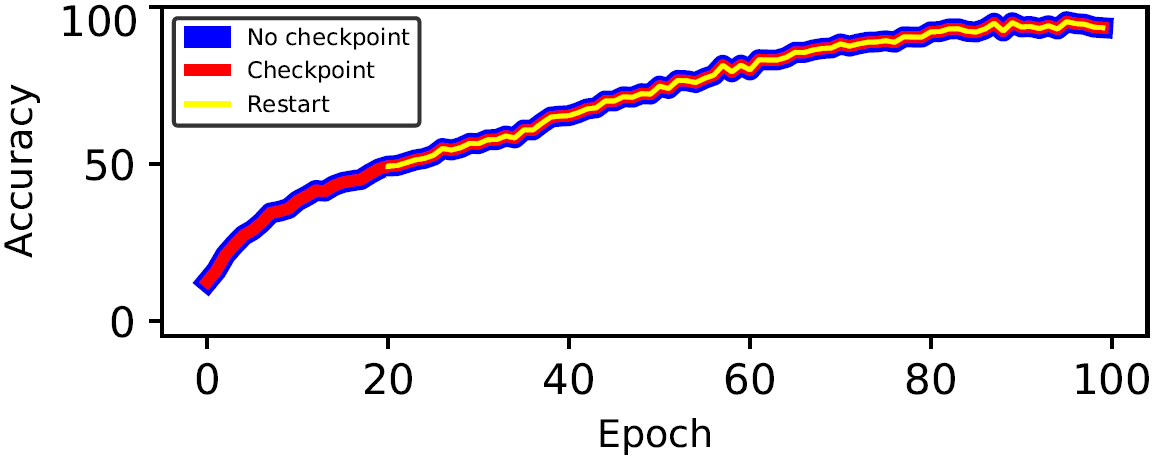}
      \caption{Accuracy.} \label{fig:pytorch_acc}
    \end{subfigure}
     \hfill
    \begin{subfigure}{.49\hsize}
      \includegraphics[width=\hsize]{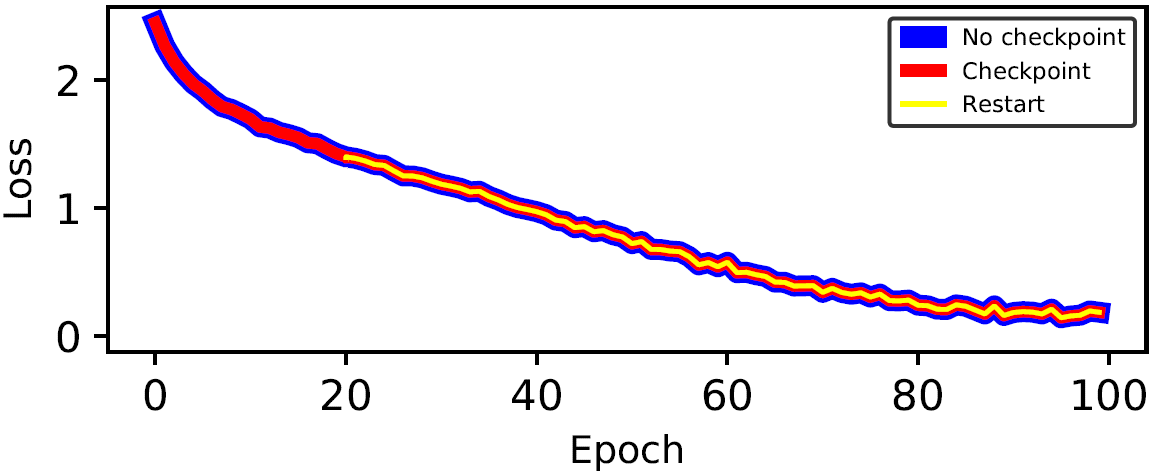}
      \caption{Loss.} \label{fig:pytorch_loss}
    \end{subfigure}
    \caption{Deterministic PyTorch distributed training. In blue the accuracy and loss are shown without performing a checkpoint, in red the accuracy and loss doing a checkpoint every 5 epochs and in yellow the restart from a checkpoint in epoch 20.}
    \label{fig:pytorch_acc_loss}
  \end{center}
\end{figure*}

\begin{figure*}[t]
  \begin{center}
    \begin{subfigure}{.32\hsize}
      \includegraphics[width=\hsize]{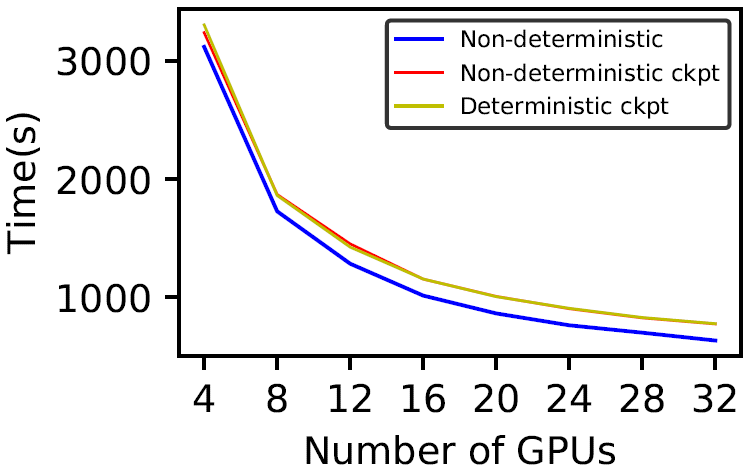}
      \caption{Chainer.}
      \label{fig:chainer_perform}
    \end{subfigure}
    \hfill
    \begin{subfigure}{.32\hsize}
      \includegraphics[width=\hsize]{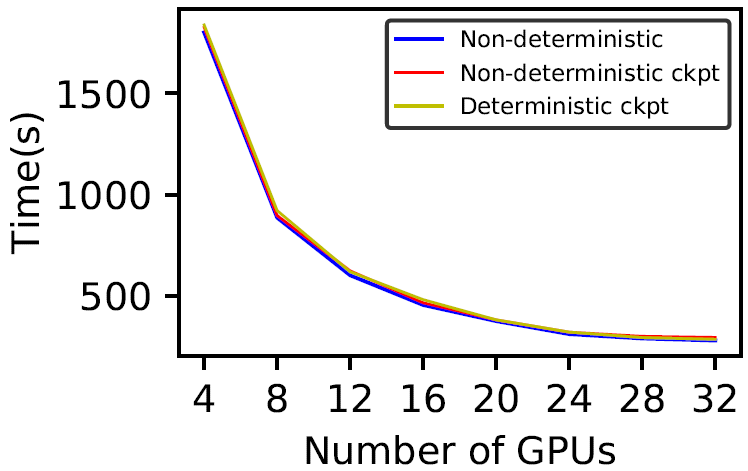}
      \caption{PyTorch.} \label{fig:Horovod_perform}
    \end{subfigure}
     %\hfill
    %\begin{subfigure}{.24\hsize}
    %  \includegraphics[width=\hsize]{figures/pytorch_performance.png}
    %  \caption{PyTorch DDP.} \label{fig:pytorch_perform}
    %\end{subfigure}
     \hfill
    \begin{subfigure}{.32\hsize}
      \includegraphics[width=\hsize]{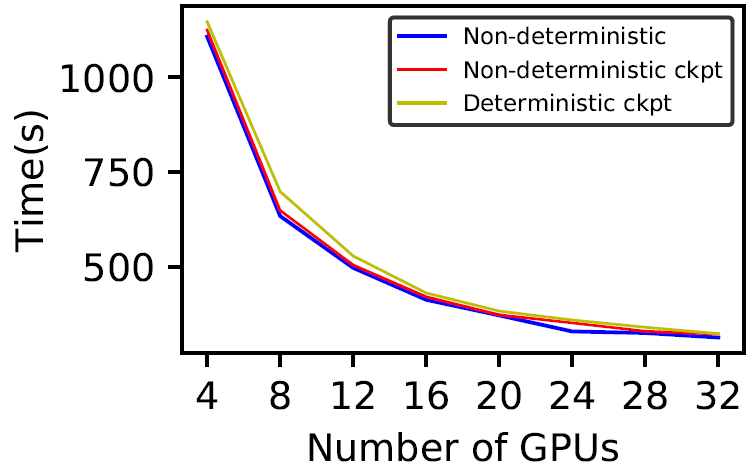}
      \caption{TensorFlow.} \label{fig:tensorflow_perform}
    \end{subfigure}
    \caption{Performance of distributed training  by each of the DL frameworks. In each figure the execution time of a non-deterministic training without checkpoint, a non-deterministic training with checkpoint, and a deterministic training with checkpoint are compared.}
    \label{fig:chainer_pytorch_horovod_perform}
  \end{center}
\end{figure*}

\subsection{Checkpointing at Scale}\label{sec:checkpoint_scale}

This set of experiments was performed on the ABCI supercomputer. Our goal is to evaulate checkpointing overhead at scale. We train two models, VGG16 and ResNet50 on the Cifar10 dataset with checkpointing and without checkpointing. For each framework, we set checkpointing at every 5 epochs. This is not ideal, and checkpointing time is determined by several factors such as length of epoch or state of the model during training (i.e. checkpoint only when there is an increase in validation accuracy). For longer training such as ImageNet, one would checkpoint at every epoch.   

Table \ref{tab:abci_times} shows the results of training on ResNet50 and VGG16 with and without checkpointing for Chainer, PyTorch and TensorFlow. As mentioned in section \ref{sec:methodology}, VGG16 and ResNet model have different properties and this is seen in the results. It is worth noting that as expected, all frameworks scale well as we increase the number of GPUs. There is a notable difference in the training times for each framework with Tensorflow performing slightly faster than other frameworks when training at small scale but performing worse when running at large scale. This shows that TensorFlow is a framework that has been highly optimized to run in single, node executions, but it has not been optimized to run in a large scale distributed system. Chainer, on the other hand, is a framework that was conceived from the beginning to scale and this can be observed in the results.  

However, Chainer has the highest checkpointing overhead and this overhead increases as the number of GPUs increases (ResNet model). This is due to the fact that in Chainer, checkpointing is a sequential operation performed by only one process while other processes wait. As such, the time to checkpoint will remain the same whether we use 1 GPU or 256 GPUs. When training the same model with checkpointing, the time for checkpointing is not large enough to be noticed when the number of GPUs is low. However at large scale, especially at 256 GPUs, the total time is double or more than the training time without checkpointing. On the contrary, Tensorflow performs slightly better than PyTorch and has the lowest checkpoint overhead. It is clear from these results that Chainer and TensorFlow have been optimized in very different ways, and this is demonstrated in these results. 

As mentioned early, VGG16 is a deep model with more parameters. Unlike ResNET, VGG16 has no skip connections and generally the model is known for being notoriously slow to train. However, since we are training the model on the Cifar10 dataset (smaller images), the training time is not too long. In this case, when training without checkpointing,  Chainer shows the best performance and PyTorch is noticeably slow in comparison. As expected, VGG16 on average takes longer to checkpoint than ResNet. This is also related to the checkpoint sizes in Table \ref{tbl:check_sizes}.  Chainer has a significant overhead that affects even the scaling behaviour. Even with just 4 GPUs, checkpoint overhead for Chainer is significantly high.  Though not the same as Chainer, a significant checkpointing overhead is also observed in PyTorch especially as the number of GPUs increases. In the case of VGG16 as well, TensorFlow has the lowest checkpointing overhead.

%---------------------------------Deterministic Checkpointing-----------------------% 
\subsection{Deterministic Checkpointing}\label{sec:deterministic_results}

As mentioned in Section~\ref{sec:methodology}, we made the necessary modifications to the DL frameworks to obtain deterministic results. Figure~\ref{fig:pytorch_acc_loss} shows accuracy and loss during training. Three types of deterministic results are shown. Figure~\ref{fig:pytorch_acc} compares $i)$ the accuracy of training without checkpointing, $ii)$ the accuracy of training with checkpointing, and $iii)$ the accuracy of training after restart from epoch 20. The same description applies to Figure~\ref{fig:pytorch_loss} but in regards to the loss.

With PyTorch it was possible to carry out a deterministic distributed training, both in a complete training cycle (100 epochs) and after restarting from a checkpoint (deterministic restart) which validates the operation of the checkpoint mechanism.  We modified the checkpoint process to include data from the optimizer. Figures~\ref{fig:pytorch_acc} and~\ref{fig:pytorch_loss} show that from epoch 20 onwards (yellow line) values of accuracy and loss are exactly the same. That way, we eliminate the natural randomness generated by training results and the randomness that increases in distributed training, in which the synchronization of processes is not deterministic.
%Now we can eliminate the uncertainty generated by the training results (always with a random nature) and the uncertainty that increases in distributed training in which the synchronization of processes can raise reasonable doubts about the veracity of the results.  We can conclude that the checkpoint mechanisms work correctly.

In the case of Chainer and TensorFLow, the accuracy and loss results were replicated, obtaining a deterministic behavior in a complete \emph{failure-free} training run. However, we could not achieve a full deterministic restart.  With Chainer, it was possible to generate an almost identical behavior pattern. Table~\ref{tbl:chainer_acc_loss} shows the accuracy and loss values every 10 epochs for a failure-free execution and compares it with an execution restarted from epoch 20. It can be seen that the values vary (e.g. epoch 20, accuracy is 0.740589 and after restart is 0.740552) and we do not get a deterministic restart.

Not being able to obtain the same deterministic results in Chainer and TensorFlow after restart is not related to a malfunction of the checkpoint mechanism. Rather, it is the result of not having enough and reliable elements in the framework to modify the non-deterministic state, the difficulty that comes with a deeper manipulation of the framework and the impossibility of manipulating optimised third-party libraries that are used especially in distributed training.

%Table~\ref{tbl:times} shows detailed information on the execution of deterministic distributed training up to 32 GPUs. Each of the rows show us the times with the execution of checkpoint and without the execution of checkpoint. The calculated execution times for each DL framework are divided into 3 categories,$i)$ Normal training (Normal): Non-deterministic execution time of the DL application, $ii)$ Checkpoint (Ckpt): DL application execution time, performing a checkpoint every 10 epochs, and $iii)$ Overhead ($\Omega$): It is the overhead of a training that executes checkpoints regards to a workout without a checkpoint.

%Figures~\ref{fig:performance_DL_frameworks} and \ref{fig:chainer_pytorch_horovod_perform} were originated from information in this table. In the figure~\ref{fig:performance_DL_frameworks} it is possible to compare the scalability when performing checkpoints in deterministic and non-deterministic training in the 3 DL frameworks (PyTorch Horovod is also included). In both figures it can be seen that TensorFlow has a lower performance and does not scale well when increasing the number of GPUs as with the other DL frameworks.

Previously, the performance of non-deterministic training with and without a checkpoint was analysed. Now, we analyse the performance of deterministic training with the execution of checkpoint. Figure~\ref{fig:chainer_pytorch_horovod_perform} shows the performance of each of the DL frameworks when running distributed training. Figures show the time it takes to complete a full training (100 epochs) according to the number of GPUs.  With these, it is possible to compare the performance of non-deterministic training without checkpoint, non-deterministic training with checkpoint, and deterministic training with checkpoint for each DL framework.

In the 3 figures, the execution of deterministic training with checkpoint follows the same behaviour as non-deterministic training with checkpoint, presenting an almost imperceptible difference in performance (red line and yellow line). If we calculate the average time difference between non-deterministic checkpoint and deterministic checkpoint, for the entire range, we get 14.1 seconds for Chainer, 19.8 seconds for PyTorch, and 16.7 for TensorFlow.  These time differences are relatively low, if we take into account that the calculation was made based on the training times carried out with amounts of GPUs ranging from 4 GPUs to 32 GPUs. 
%($AVG(|Non\_det\_ckpt\_time_{gpu_i}-deterministic\_ckpt\_time_{gpu_i}|)$)

Time variations  can be attributed to latencies generated by the parallel file systems or the network.  Of course, the completion time of a checkpointed training increases depending on the number of checkpoints that need to be performed during the training. The results show us that deterministic behavior does not influence the performance of distributed training with the execution of checkpoint.

\begin{table}[t]

\begin{minipage}{.45\textwidth}
\centering
\footnotesize
\caption{Accuracy and loss values of a distributed training with Chainer.}
\setlength{\tabcolsep}{7.5 pt}
\begin{tabular}{c|cc|cc}\hline
\toprule[0.5pt]
\midrule[0.3pt]
 & \multicolumn{2}{c|}{\textbf{Accuracy}} & \multicolumn{2}{c}{\textbf{Loss}}  \\\hline
\textbf{Epoch}  & \textbf{Training} & \textbf{After restart}  &\textbf{Training} & \textbf{After restart}  \\\hline
10	&	0.248935	& - 	    & 2.226590 & -	 			\\%\hline
20	&   0.740589 & 0.740552 	& 0.743844 & 0.75368		\\%\hline
30	&	0.789240	&  0.789595	& 0.615287  & 0.614189		\\%\hline
40	&	0.810369	& 0.809659	& 0.549811 & 0.550377	    \\%\hline
50	&	0.832741	& 0.833629	& 0.478858 & 0.478444			\\%\hline
60	& 0.854403		& 0.855291	& 0.430454 & 0.429604			\\%\hline
70	& 0.861683		& 0.862216	& 0.389529 & 0.386645			\\%\hline
80	& 0.878374		& 0.876953	& 0.353764 & 0.354646	 		\\%\hline
90	& 0.887429		& 0.888139	& 0.323122 & 0.319450		\\%\hline
100	& 0.902344		& 0.900923	& 0.283951 & 0.286143		\\\hline

\toprule[0.5pt]
\end{tabular}

\label{tbl:chainer_acc_loss}
%\todo{Why is this table in methodology section }
\end{minipage}
\end{table}

\section{Discussion}\label{sec:discussion}

Checkpointing is an integral part of DL workloads in HPC. Our study has unveiled several insights that may help both framework developers and users when training DL models: 

\begin{itemize}

    %Such frameworks would have to be heavily modified to facilitate checkpointing in a data parallelism fashion. 
    
    %There is already some promising work done along this line by \cite{! }. However the work is not conclusive since no analysis was performed on the cost of  reconstructing the model and restarting. 

    \item \textbf{\emph{Checkpoint file format}}: Each DL framework has its own file format for saving checkpoints. This is expected as the checkpoint mechanism is not standard. There are clear differences between file sizes due to the choice of file formats and compression mechanisms. For instance, Chainer is able to significantly compress checkpoint files. To facilitate using existing checkpointing libraries for DL workloads, the file format should be addressed. For instance, a framework agnostic checkpointing library should be able to checkpoint in a file format that frameworks understand and are able to resume from.

    \item \textbf{\emph{Checkpoint implementations:}}
        All DL frameworks explored in this study have the ability to checkpoint during training in a native way.  Unfortunately, DL frameworks have evolved to adapt to new HPC systems that allow distributed training without taking checkpointing into account. There are checkpointing libraries specifically designed for HPC applications~\cite{fti,veloc}. Those libraries are optimized for better performance under distributed processing conditions. There is already an effort to adapt a checkpoint library developed for HPC environments to DL applications~\cite{nicolae}. The idea is to offer DL applications advanced features such as asynchronous multi-level checkpointing, hiding the complexity of heterogeneous storage, or providing efficient serialization on local storage, which are some of the fundamental elements to obtain the maximum performance. All these elements are not present in the current DL frameworks.

%Currently, DL framework developers are probably more interested in optimizing, debugging, and standardizing the mechanisms that drive distributed training.  At some point they will have to turn their gaze to the checkpoint processes that will take on more relevance as training times increase and checkpoint file sizes grow, punishing the scaling.  With this scenario, we can think that the implementation of already developed checkpoint libraries is a good choice to solve the possible drawbacks of checkpointing in DL distributed applications.  However, efforts to carry out this type of implementations so far are few and it is not known if the developers of DL frameworks are interested in taking the ideas from the existing checkpointing libraries and adding them to their frameworks or developing native implementations with the same features needed to take advantage of HPC systems.

\begin{table}[t]
\caption{\revision{Checkpointing Features of DL Frameworks.}}
\setlength{\tabcolsep}{7.6 pt}
\begin{tabular}{l|c|c|c}
\toprule[0.5pt]
\midrule[0.3pt]
\centering
& PyTorch & Tensorflow & Chainer \\ \hline
Checkpoint Implementation    & \checkmark       & \checkmark         & \checkmark  \\ \hline
Multi-Node Checkpointing     & -       & -         & \checkmark      \\ \hline
Deterministic Training       & \checkmark       & \checkmark          & \checkmark \\ \hline
Model Parallel Checkpointing & -       & -          & -       \\ \hline
\toprule[0.5pt]
\end{tabular}

\label{tab:summary}
\end{table}

    \item \textbf{\emph{Checkpoint scalability:}} Increasing the number of GPUs significantly improves training time of DL models. However, using multiple nodes does not improve the checkpointing time as only one node is in charge of checkpointing. Though Chainer has some sort of multi-node checkpoint implementation, our study showed that this simply produces redundant copies of the application's state and does not improve the time to checkpoint. There is therefore an opportunity for external libraries typically used in HPC to improve checkpointing in DL workloads. 

 \item \textbf{\emph{Deterministic behavior in DNN: }} Determinism in DL models is a secondary consideration. That might be expected since ANNs have a random nature that seeks to model the behavior of neurons in a biological brain. However, the randomness of the results can raise reasonable doubts about their veracity as they are not reproducible, especially when DL models grow in complexity and are susceptible to failures. This is particularly important for critical-mission applications such as self-driving cars, among others. We realised that the options offered by DL frameworks to generate 100\% deterministic results in training and when restarting a training (deterministic restart) are not always reliable to generate reproducible results. 
 %In some cases the developers themselves do not ensure complete deterministic behaviors between versions and/or platforms (hardware, software).  
 It is imperative for DL frameworks to provide clear and easy-to-implement mechanisms that allow the reproducibility of results and therefore research environments that allow validation of the experiments.

    \item \textbf{\emph{Data parallel vs model parallel:}} Though this study primarily focused on data parallelism, it is also important to consider model parallelism especially now that it is becoming popular due to increase in model sizes. In the data parallelism paradigm, there is a replica of the model on each GPU. In model parallelism, the model is split between processes and each process trains a part of the model. Existing checkpointing implementations in DL frameworks do not do partial checkpoints. The challenge is that all processes have a replica of the model and all perform an all-reduce operation at the end of an iteration. To scale checkpointing up, the model has to be broken up, so that each process checkpoints a small part of it. 
    
\end{itemize}

\revision{A summary of the findings discussed in this section and the checkpointing properties of the different frameworks is given in Table~\ref{tab:summary}.}

\section{Conclusion}\label{sec:conclusion}

Checkpointing is a fundamental component when training DL models in HPC due to the lengthy training times and high probability of hardware faults in HPC systems. In this paper, we have evaluated through a series of experiments the checkpointing implementations of three state-of-the-art DL frameworks.  We have evaluated factors such as computational cost of checkpointing, file sizes and formats, the effect of scale and deterministic behaviour. All frameworks have a form of checkpointing support that is considered sufficient. However, our evaluation has shown that at scale, this checkpoint implementations come with a significant overhead as many GPUs remain idle during checkpointing, particularly for Chainer and Pytorch. The file size of the checkpoint changes significantly with the model in Chainer unlike Tensorflow and Pytorch. The insights provided in this paper can help users and framework developers to guide future developments of advanced checkpointing libraries for DL workloads in HPC. \revision{As future work we are interested in expanding the parameters of the experiments including more DL models with resilience analysis. Because these models directly affect the size and performance of the checkpoint process. In addition, we are interested in research on the implementation of more efficient checkpoint mechanisms based on consolidated libraries that are used in the HPC area.}

%\section*{Acknowledgment}

\footnotesize

\vspace{1mm}
\noindent
\textbf{Acknowledgments}. This project received support and funding from the EuroLab4HPC project. European Union Horizon 2020 Framework Programme (\textbf{H2020-EU.1.2.2. - FET Proactive}) under grant agreement number \textbf{800962}. 

The project that gave rise to these results received the support of a fellowship from the ``la Caixa" Foundation (ID \textbf{ 100010434}). The fellowship code is \textbf{LCF/BQ/DI17/11620059}. This project has received funding from the European Union's Horizon 2020 research and innovation programme under the Marie Skłodowska-Curie grant agreement No.\textbf{713673}.

\revision{ Rosa M Badia  has also been supported by the Spanish Government through contracts SEV2015-0493 and TIN2015-65316-P, and by Generalitat de Catalunya through contract 2014-SGR-1051.}

\bibliographystyle{IEEEtran}
\bibliography{IEEEabrv,bib}

\end{document}